\documentclass[aps,pra,twocolumn,superscriptaddress,longbibliography, nofootinbib]{revtex4-2}
\usepackage[utf8]{inputenc}
\usepackage{amsmath}
\usepackage{amssymb}
\usepackage{amsthm}
\usepackage{dsfont}
\usepackage{bm}
\usepackage{appendix}
\usepackage{comment}
\usepackage[unicode=true,pdfusetitle,
bookmarks=true,bookmarksnumbered=false,bookmarksopen=false,
 breaklinks=true,pdfborder={0 0 0},pdfborderstyle={},backref=false,colorlinks=true]
 {hyperref}

\newcommand\R{{\mathbb R}}

\newcommand\tr{{\rm Tr}}

\newcommand\dps{\displaystyle }

\usepackage{tcolorbox}

\begin{document}
\title{New perspectives on Density-Matrix Embedding Theory}
\author{Alicia Negre}
\affiliation{CERMICS, Ecole des Ponts - Institut Polytechnique de Paris and Inria Saclay, France}
\author{Fabian Faulstich}
\affiliation{Department of Mathematics, Rensselaer Polytechnic Institute, Troy, New York, USA}
\author{Raehyun Kim}
\affiliation{Department of Mathematics, University of California, Berkeley}\author{Thomas Ayral}
\affiliation{Eviden Quantum Lab, 78340 Les Clayes-sous-Bois, France}
\author{Lin Lin}
\affiliation{Department of Mathematics, University of California, Berkeley}
\affiliation{Lawrence Berkeley National Laboratory}
\author{Eric Canc\`es}
\affiliation{CERMICS, Ecole des Ponts - Institut Polytechnique de Paris and Inria Saclay, France}

\begin{abstract}
Quantum embedding methods enable the study of large, strongly correlated quantum systems by (usually self-consistent) decomposition into computationally manageable subproblems, in the spirit of divide-and-conquer methods. Among these, Density Matrix Embedding Theory (DMET) is an efficient approach that enforces self-consistency at the level of one-particle reduced density matrices (1-RDMs), facilitating applications across diverse quantum systems. However, conventional DMET is constrained by the requirement that the global 1-RDM (low-level descriptor) be an orthogonal projector, limiting flexibility in bath construction and potentially impeding accuracy in strongly correlated regimes. In this work, we introduce a generalized DMET framework in which the low-level descriptor can be an arbitrary 1-RDM and the bath construction is based on optimizing a quantitative criterion related to the maximal disentanglement between different fragments. This yields an alternative yet controllable bath space construction for generic 1-RDMs, lifting a key limitation of conventional DMET. We demonstrate its consistency with conventional DMET in appropriate limits and exploring its implications for bath construction, downfolding (impurity Hamiltonian construction), low-level solvers, and adaptive fragmentation. We expect that this more flexible framework, which leads to several new variants of DMET, can improve the robustness and accuracy of DMET.
\end{abstract}

\maketitle

\section{Introduction}
\label{sec:introduction}

Quantum embedding methods \cite{Zgid2011, Sun2016a, maier2005quantum} are scalable protocols to compute properties of large interacting quantum systems, including lattice models, quantum chemical systems, and correlated materials. These methods employ a self-consistent construction of smaller and/or less complex many-body problems by leveraging an approximate, computationally inexpensive (``low-level'') descriptor of the original system, which are then solved with highly accurate (``high-level") methods.

Different embedding approaches impose distinct self-consistency conditions between the high-level and low-level descriptors. For instance, dynamical mean-field theory (DMFT) \cite{georges1996dynamical} enforces consistency of (time-dependent) one-particle Green’s functions, while density-matrix embedding theory (DMET) \cite{DMET2012} ensures matching one-particle reduced density matrices (1-RDM). Other methods, such as density embedding theory (DET), impose self-consistency at the level of densities. These choices fundamentally determine the structure and properties of the resulting surrogate many-body problems \cite{Ayral2017a}.
Each problem describes a small subsystem, called a fragment, coupled with a bath that mimics the effects of the full system at the requisite level of self-consistency.
Requiring self-consistency at the level of more complex objects (e.g. time-dependent objects, as opposed to static objects) usually leads to more intricate bath constructions.
For instance, DMFT in principle requires infinite baths (see~\cite{CKPR2024} for a mathematical proof in the framework of iterative perturbation theory), while conventional DMET requires a finite bath whose size equals that of the fragment (in nondegenerate cases).

The 1-RDM-level self-consistency condition of DMET~\cite{DMET2012, DMET2013, tsuchimochi2015density, bulik2014density, wouters2016practical, cui2019efficient,sun2020finite,cui2020ground} enables numerically less expensive implementations compared to DMFT.
Over the years, this has allowed DMET to be applied to a wide range of systems such as Hubbard models~\cite{DMET2012, bulik2014density, chen2014intermediate, boxiao2016, Zheng2017, zheng2017stripe, welborn2016bootstrap,senjean2018site,senjean2019projected}, quantum spin models~\cite{Fan15, gunst2017block,ricke2017performance}, and a number of strongly correlated molecular and periodic systems~\cite{DMET2013, wouters2016practical,cui2020ground,nusspickel2022systematic,bulik2014electron,pham2018can,hermes2019multiconfigurational,tran2019using,ye2018incremental,ye2019bootstrap,ye2019atom,ye2020bootstrap,ye2021accurate,tran2020bootstrap,meitei2023periodic,mitra2022periodic,mitra2021excited,ai2022efficient}. 
Recently, DMET variants have even been considered for execution on quantum computers~\cite{Rubin2016,Besserve2021,vorwerk2022quantum,cao2022ab,shajan2024towards}.

Various approaches have been explored to define the bath space in embedding methods, including different choices of low-level theories~\cite{fertitta2018rigorous, nusspickel2020efficient, nusspickel2020frequency, Sekaran2022}, alternative constructions and solutions for the Hamiltonian of the interacting fragment-plus-bath system~\cite{nusspickel2022effective, potthoff2001two, lu2019natural, ganahl2015efficient, scott2021extending, Sekaran2021, yalouz2022quantum, Marecat2023a}, and modifications to self-consistency conditions~\cite{wu2019projected, wu2020enhancing, faulstich2022pure}. A fundamental limitation of conventional DMET is its reliance on a low-level descriptor---the global 1-RDM---being an idempotent matrix, i.e., an orthogonal projector. This idempotency is crucial for constructing the bath and formulating the surrogate many-body problems in conventional DMET. To address this constraint, EwDMET~\cite{fertitta2018rigorous, Fertitta2019}---a DMET variant that seeks to incorporate impurity-environment correlations beyond conventional DMET---introduces auxiliary modes and matches ``energy-weighted" density matrices, thereby relaxing the self-consistency requirement while extending the applicability of the method. 
Another embedding-type approach that targets strongly-correlated electronic systems is the rotationally-invariant slave-boson theory (RISB)~\cite{fresard1992unified,lechermann2007rotationally,lanata2017slave,lee2019rotationally}, which is equivalent to the multi-orbital Gutzwiller approximation at the mean-field level~\cite{kotliar1986new,bunemann2007equivalence,lanata2008fermi}.
In its own way, it indirectly addresses the idempotency issue by supplementing the correlation potential with a renormalized hopping factor as an additional degree of freedom to reach self-consistency \cite{Ayral2017a}.

In this article, we present a generalized formulation of DMET that allows for a flexible bath construction from arbitrary global 1-RDMs. Our construction avoids the use of orthogonal projectors in the bath construction while keeping the bath space size controlled and allows any \textit{a priori} chosen bath size. The bath construction is based on optimization of a disentanglement criterion (vide infra) providing an alternative and systematic approach to the bath construction of specified sizes---either user-defined or adaptively chosen. The optimization perspective moreover provides an {\it a posterior} evaluation of the fragmentation, namely, if the 1-RDM segments cannot be sufficiently disentangled, the method is likely to yield inaccurate results. This more flexible construction may alleviate some limitations of conventional DMET, such as a certain rigidity leading to duality gaps \cite{faulstich2022pure,wu2020enhancing} and $N$-representability issues~\cite{CFKLL2025}.

This article is structured as follows: In Section~\ref{sec:notation}, we introduce the physical model and establish the notation used throughout the paper. Section~\ref{sec:DMET} defines the five key components of Density Matrix Embedding Theory (DMET): (i) fragmentation, (ii) bath construction, (iii) downfolding (impurity Hamiltonian construction), (iv) charge equilibration, and (v) low-level solver. 
In Section~\ref{sec:bath_construction}, we introduce a new bath construction method that applies to any admissible global 1-RDM $D$ and allows for an arbitrary bath dimension. This construction aims to maximally disentangle---in a sense specified later---the fragment-plus-bath subsystem from the rest of the system. Notably, the proposed approach reduces to the conventional DMET bath construction when $D$ is an orthogonal projector and the bath size equals the fragment size.
Section~\ref{sec:downfolding} demonstrates that this generalized bath construction leads to a canonical downfolding procedure, naturally defining impurity Hamiltonians. Again, our approach coincides with conventional DMET when $D$ is an orthogonal projector and the fragment and bath have equal dimensions. Section~\ref{sec:low-level} discusses low-level solvers and their connection to (Reduced) Density-Matrix Functional Theory ((R)DMFT), using this notation to distinguish it from Dynamical Mean-Field Theory (DMFT). Section~\ref{sec:fragmentation} explores the potential for adaptive fragmentation, where the fragment decomposition is not chosen \emph{a priori} but emerges iteratively through optimization over flag varieties. This approach leads to several new DMET variants, which will be tested and compared in future work.

\section{Physical model and notation}
\label{sec:notation}

\begin{figure}
    \centering
    \includegraphics[width=1.\columnwidth]{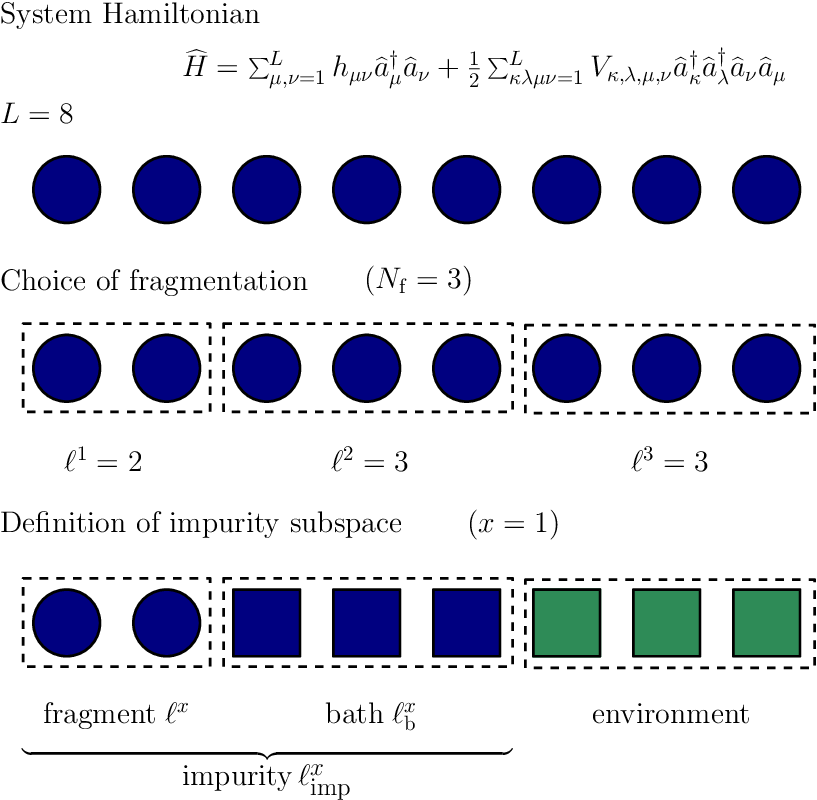}
    \caption{Definition of the various subspaces used in DMET. $L$ is the total number of orbitals, $N$ the total number of electrons, $N_\mathrm{f}$ the number of fragments, $\ell^x$ the size of the $x$th fragment and $\ell^x_\mathrm{b}$ the size of its bath (in conventional DMET, $\ell^x_\mathrm{b} = \ell^x$). The different shapes used for the orbitals specify the bases in which they are represented.}
    \label{fig:dmet_defs}
\end{figure}

We consider a system of $N$ electrons on $L_{\rm s}$ space orbitals (or sites) described by a Hamiltonian including one- and two-body interaction terms. In particular, we have electronic Hamiltonians and fermionic Hubbard-like models in mind. In DMET, the global system is partitioned into several subsystems called fragments which, in conventional DMET, are fixed and each fragment has a corresponding impurity which, in conventional DMET, is twice as big as the fragments. The impurities are auxiliary systems that approximate the interactions between the individual fragments with the remaining system via a bath obtained by compressing said remaining system. In conventional DMET, this interaction is obtained from the 1-RDM, which is updated through the procedure, hence, the bath and therewith the impurity may change during the computational procedure. These different systems are illustrated in Fig.~\ref{fig:dmet_defs}.

Although non-physical, one may interpret the fragments as open systems exchanging particles with their baths. However, only the solutions on the fragments are used to form a global descriptor in DMET. Consequently, the global particle number which is given as the sum of the particle numbers per fragment is in general not fixed, and the mean value of this observable is in general not an integer. For these reasons, the mathematical formulation of DMET involves several one-particle state spaces (one for the whole system, one for each fragment, and one for each impurity) and the corresponding Fock spaces. Additionally, since each impurity problem in the DMET algorithm employs a different basis for the one-particle state space, adapted to the impurity-plus-environment decomposition at each macro-iteration, it is more natural to formulate the theory in terms of operators rather than matrices.

\medskip

To simplify the notation, we work in the non-relativistic setting and assume that there is no external magnetic field acting on the spins, which allows us to work with real vector spaces. We denote by $\mathcal H$ the one-particle state space of the whole system, i.e. the real linear space spanned by the $L=2L_{\rm s}$ spin-orbitals and by 
\begin{equation} \label{eq:Fock-space}
\mathcal H_n:=\bigwedge^n \mathcal H \quad \mbox{and} \quad {\rm Fock}(\mathcal H):= \bigoplus_{n=0}^{L} \mathcal H_n
\end{equation}
 the $n$-particle state space (the antisymmetrized tensor product of $n\in[\![L]\!]$ copies of the one-particle state space) and the Fock space of the whole system, respectively. Here and below $[\![k]\!]$ denotes the set of integers $\{1,2,\cdots,k\}$. We denote by 
$$
\widehat H =\widehat H_{\rm 1-body} + \widehat H_{\rm 2-body} 
$$
the Hamiltonian of the whole system in the second quantization formalism. The operator $\widehat H$ is symmetric on ${\rm Fock}(\mathcal H)$ and particle-number conserving (i.e. it commutes with the particle-number operator), which implies that it is block-diagonal in the decomposition \eqref{eq:Fock-space} of the Fock space into $n$-particle sectors. Likewise, if $\mathcal X$ is a linear subspace of $\mathcal H$ (typically $\mathcal X$ will be the state space of some impurity problem), we denote by 
$$
{\rm Fock}(\mathcal X):= \bigoplus_{n=0}^{{\rm dim}(\mathcal X)} \bigwedge^n \mathcal X
$$
the Fock space associated by $\mathcal X$.

\medskip

In conventional DMET, the global state of the system is described at each iteration by a Slater-type 1-RDM, i.e. a rank-$N$ orthogonal projector corresponding to a unique $N$-body state represented by a Slater determinant. The set of such 1-RDMs is
\begin{align*}
{\mathcal D}_{\rm Slater}:&=\{ D \in {\mathcal S}(\mathcal H) \; | \; D^2=D, \; \tr(D)=N\}
\end{align*}
where ${\mathcal S}(\mathcal H)$ and $\tr(D)$ denote the space of symmetric operators on $\mathcal H$ and the trace of $D$, respectively. In the variants of DMET we consider, we will work with 1-RDMs in the set 
$$
{\mathcal D}_{\rm MS}:=\{ D \in {\mathcal S}(\mathcal H) \; | \; 0 \le D \le 1, \; \tr(D)=N\},
$$
of all $N$-representable mixed-state 1-RDMs. The notation $0 \le D \le 1$ means that $D$ is between the zero operator and the identity operator in the sense of quadratic forms. This condition is equivalent to the fact that all the eigenvalues of~$D$ (i.e. the natural occupation numbers) are in the range $[0,1]$, as required by the Pauli principle. From a geometrical point of view, ${\mathcal D}_{\rm Slater}={\rm Gr}(N,\mathcal H)$, where
$$
{\rm Gr}(n,\mathcal X):= \left\{ D \in \mathcal S(\mathcal X) \; | \; D^2=D, \; \tr(D)=n \right\}
$$
is the real Grassmann manifold of the rank-$n$ orthogonal projectors in $\mathcal S(\mathcal X)$, and ${\mathcal D}_{\rm MS}$ is the convex hull of ${\mathcal D}_{\rm Slater}$. Note that the set of pure-state $N$-representable 1-RDMs does not have a simple explicit characterization.

\medskip

In order to describe the different computational steps of the DMET iteration procedure in a precise way, it is convenient to work in specific basis sets that depend on the iteration step. For instance to explain the construction of the bath orbitals for a given fragment of size~$\ell$, it is convenient to work in an orthogonal basis of~$\mathcal H$ in which the first $\ell$ basis functions form an orthonormal basis of the fragment subspace. It is important to note that in the following, the basis set is not fixed and will change from section to section. 

Consider an orthonormal basis $(\chi_1,\cdots,\chi_L)$ of $\mathcal H$ and denote by $\widehat a_\mu^\dagger$ and $\widehat a_\mu$ the creation and annihilation operator of the spin-orbital~$\chi_\mu$, respectively. Since we consider fermions, these operators satisfy the canonical anticommutation relations
$$
\{\widehat a_\mu,\widehat a_\nu\}=0, \quad \{\widehat a_\mu^\dagger,\widehat a_\nu^\dagger\}=0, \quad \{\widehat a_\mu,\widehat a_\nu^\dagger\}=\delta_{\mu\nu},
$$
and it holds
\begin{align*}
\widehat H &=\widehat H_{\rm 1-body} + \widehat H_{\rm 2-body} \\
&=\sum_{\mu,\nu=1}^L h_{\mu\nu} \widehat a_\mu^\dagger \widehat a_\nu + \frac 12 \sum_{\kappa,\lambda,\mu,\nu=1}^L V_{\kappa\lambda\mu\nu} \widehat a_\kappa^\dagger  \widehat a_\lambda^\dagger \widehat a_\nu \widehat a_\mu
\end{align*}
where the matrix $h \in \R^{L \times L}$ and the fourth-order tensor $V \in \R^{L \times L \times L \times L}$ satisfy the 8-fold symmetry:
$$
h_{\nu\mu}=h_{\mu\nu}, \quad V_{\kappa\lambda\mu\nu} =V_{\mu\nu\kappa\lambda}= V_{\lambda \kappa \nu\mu} = V_{\nu \mu \lambda\kappa}
$$
which make $\widehat H$ a real symmetric operator. 
Note that we are using the physicist notation for the two-body interaction 
$$
V_{\kappa\lambda\mu\nu}
=
\int
\chi_\kappa(\mathbf x_1) \chi_\lambda(\mathbf x_2) V(|\mathbf r_1-\mathbf r_2|)\chi_\mu(\mathbf x_1)\chi_\nu(\mathbf x_2) \, d\mathbf x_1 \, d\mathbf x_2, 
$$
where $\mathbf{x}=(\mathbf{r}, \sigma)$ is the space-spin coordinate. 
In conventional DMET, the low-level problem is formulated as a constrained Hartree-Fock problem. Using the orthonormal basis $(\chi_1,\cdots,\chi_L)$ to identify $\mathcal H$ with $\R^L$, 1-RDMs are then represented by real-symmetric matrices and the Hartree-Fock energy functional $E^{\rm HF}:\R^{L \times L}_{\rm sym} \to \R$ is defined by
$$
E^{\rm HF}(D) = \tr(hD)+\frac 12 \tr(J(D)D) - \frac 12 \tr(K(D)D),
$$
where the second and third terms in the right-hand side are the Hartree and exchange terms, respectively. The operators $J:\R^{L \times L}_{\rm sym} \to \R^{L \times L}_{\rm sum}$ and $K: \R^{L \times L}_{\rm sym} \to \R^{L \times L}_{\rm sym}$ are defined as
\begin{align*}
[J(D)]_{\mu\nu} :&= \sum_{\kappa,\lambda=1}^L V_{\mu\kappa\nu\lambda} D_{\kappa\lambda}, \\
[K(D)]_{\mu\nu} :&=  \sum_{\kappa,\lambda=1}^L V_{\mu\kappa\lambda\nu} D_{\kappa\lambda} . 
\end{align*}

\section{Basic ingredients of Density Matrix Embedding Theories}
\label{sec:DMET}

\noindent
The DMET algorithm can be understood as an iterative procedure comprising two alternating steps:

\begin{itemize}
    \item \textbf{Construction of local many-body problems:} 
    Starting from a given global 1-RDM $D_{\rm in} \in {\mathcal D}_{\rm DMET}$ (the low-level descriptor), local many-body problems are constructed. These problems are defined on lower-dimensional subspaces of the state space of the total system, reducing the computational complexity while retaining key correlations.
    \item \textbf{Reconstruction of a global 1-RDM:} Using the solutions of the local many-body (high-level) problems, a new global 1-RDM, $D_{\rm out} \in {\mathcal D}_{\rm DMET}$, is constructed. This updated density matrix is then used as input for the next iteration.
\end{itemize}

It is important to note that the term \emph{local} does not strictly refer to spatial locality. Instead, it implies that each local many-body problem is formulated on the Fock space associated with a relatively small subspace of the one-particle state space.

\medskip

In this section, we first describe the basic principles of these two steps and explain how these principles are implemented in conventional DMET. We then sketch the main differences between conventional DMET and the variants we propose.

\medskip

For conventional DMET, the set of admissible global 1-DRMs is $\mathcal D_{\rm DMET}:=\mathcal D_{\rm Slater}$, meaning that the low-level description of the state is an orthogonal projector representing an uncorrelated pure state, as in Kohn--Sham DFT, for instance.

\medskip

\subsection{Construction of local many-body problems}

The construction of local many-body problems involves three main steps. For notational simplicity, we set $D := D_{\rm in}$.

\subsubsection{Fragmentation}  
The first step is to decompose the one-particle Hilbert space $\mathcal{H}$ into a direct sum of fragment spaces:
\begin{equation} 
\label{eq:fragment_dec}
\mathcal{H} 
= 
\mathcal{H}^{1,D}_{\rm frag} \oplus \cdots \oplus \mathcal{H}^{{N_{\rm f}^D},D}_{\rm frag},
\end{equation}
where $N^D_{\rm f}$ denotes the number of fragments, and $\mathcal{H}^{x,D}_{\rm frag}$ represents the one-particle state space associated with fragment $x$. The dimension of each fragment space is given by $\ell^{x,D} := \dim(\mathcal{H}^{x,D}_{\rm frag})$. This is illustrated in Fig.~\ref{fig:dmet_defs}.

In the conventional DMET framework, the number of fragments \( N_{\rm f} \) and the fragment spaces \( \mathcal{H}^x_{\rm frag} \), are \emph{a priori} fixed, and remain unchanged throughout the iterative procedure.

We moreover assume that the individual fragments do not overlap. Extending conventional DMET to overlapping fragments is an objective of future work.

\subsubsection{Bath construction} 
For every fragment, we seek a bath space $\mathcal H_{\rm bath}^{x,D}$ of dimension ${\rm dim }(\mathcal H_{\rm bath}^{x,D})= \ell^{x,D}_{\rm b}$ such that 
$$
{\mathcal H} = \mathcal H_{\rm imp}^{x,D} \oplus \mathcal H_{\rm env}^{x,D}, 
\qquad \mathcal H_{\rm imp}^{x,D} = \mathcal H^{x,D}_{\rm frag} \oplus \mathcal H_{\rm bath}^{x,D}, 
$$
where the {\em environment} space $\mathcal H_{\rm env}^{x,D}$ is the orthogonal complement of the {\em impurity} space $\mathcal H_{\rm imp}^{x,D}$.

In conventional DMET,  $\ell^{x,D}=\ell^x$, $\mathcal H_{\rm frag}^{x,D}=\mathcal H_{\rm frag}^{x}$ and $\ell^{x,D}_{\rm b,conv}=\ell^x$ are independent of $D$, and $\mathcal H_{\rm imp,conv}^{x,D} :=  \mathcal H_{\rm frag}^{x} + D  \mathcal H_{\rm frag}^{x}$. Under a suitable non-degeneracy assumption (see Section~\ref{sec:bath_construction} for details), this construction yields ${\rm dim}\left(\mathcal H_{\rm imp,conv}^{x,D}\right)=2\ell^x$, thus ${\rm dim}\left(\mathcal H_{\rm bath,conv}^{x,D}\right)=\ell^x$ as desired.

\subsubsection{Construction of impurity Hamiltonians}

In conventional DMET, $D$ is an orthogonal projector representing a unique $N$-body Slater determinant $\Psi^D \in \dps \bigwedge^N \mathcal H$. The decomposition ${\mathcal H} =\mathcal H_{\rm imp,conv}^{x,D} \oplus \mathcal H_{\rm env,conv}^{x,D}$ yields that $\Psi^D$---in the non-degenerate case---is constructed from $\ell^x$ orthonormal spin-orbitals in $\mathcal H^{\rm imp,conv}_{x,D}$ and $(N-\ell^x)$ orthonormal spin-orbitals in $\mathcal H_{\rm env,conv}^{x,D}$, so that 
$$
\Psi^D = \Psi^{x,D}_{\rm imp}\wedge \Psi^{x,D}_{\rm env} 
$$
with
$$
\Psi^{x,D}_{\rm imp}\in \dps \bigwedge^{\ell^x} \mathcal H^{x,D}_{\rm imp,conv},
\quad \mbox{and} \quad 
\Psi^{x,D}_{\rm env} \in \bigwedge^{N-\ell^x} \mathcal H^{x,D}_{\rm env,conv}.
$$
The impurity Hamiltonian $\widehat H^{x,D}_{\rm imp,conv}$ is then defined as the unique operator on ${\rm Fock}\left(\mathcal H^{x,D}_{\rm imp,conv}\right)$ such that
\begin{equation} 
\label{eq:Himp_st}
\big\langle \Psi_{\rm imp}  | \widehat H^{x,D}_{\rm imp,conv} | \Psi_{\rm imp} \big\rangle:= \big\langle \Psi_{\rm imp} \wedge \Psi^{x,D}_{\rm env} | \widehat H |  \Psi_{\rm imp} \wedge \Psi^{x,D}_{\rm env} \big\rangle,
\end{equation}
for all
 $\Psi_{\rm imp} \in {\rm Fock}\left(\mathcal H^{x,D}_{\rm imp,conv}\right)$.

The operator $\widehat H^{x,D}_{\rm imp,conv}$ is a particle-number conserving (i.e. commuting with the impurity particle-number operator) two-body Hamiltonian, which has a simple explicit expression. See subsection~\ref{sec:downfolding} below for details.

\subsection{Reconstruction of a global 1-RDM}

Once the local many-body problems have been constructed, the next step is to solve them during the {\it charge equilibration} step. In the subsequent {\it low-level solver} step, the resulting local solutions are combined to construct a global descriptor, represented by a global 1-RDM $D_{\rm out}$.

\subsubsection{Charge equilibration}
This step aims at building the diagonal blocks of the global descriptor $D_{\rm out}$ in the fragment decomposition~\eqref{eq:fragment_dec}, i.e., 
$$
P^{x,D}_{\rm frag} D_{\rm out} \left(P^{x,D}_{\rm frag}\right)^*
=:
D^{xx,D}_{\rm out} \in \mathcal S\left(\mathcal H^{x,D}_{\rm frag}\right)
, 
$$
where $P^{x,D}_{\rm frag}: \mathcal H \to \mathcal H^{x,D}_{\rm frag}$ is the orthogonal projector from~$\mathcal H$ onto the fragment subspace $\mathcal H^{x,D}_{\rm frag}$. 

Note that although each impurity Hamiltonian $\widehat H^{x,D}_{\rm imp}$ is particle-number conserving, a grand-canonical charge equilibration procedure has to be applied to make sure that the global descriptor $D_{\rm out}$ has the correct particle number. Indeed, while the fragments themselves do not overlap, the impurities—each consisting of a fragment and its corresponding bath—overlap; hence, fixing the number of electrons in each impurity by a microcanonical approach is infeasible. To guarantee that the overall system has $N$ electrons, it is necessary to monitor the number of electrons in each fragment. We thus introduce two number operators on the Fock space ${\rm Fock}\left(\mathcal H^{x,D}_{\rm imp}\right)$, the standard particle-number operator $\widehat N^{x,D}_{\rm imp}$, which counts the total number of particles in the impurity, and the operator $\widehat N^{x,D}_{\rm frag}$, which counts the number of particles in the fragment. Consider an impurity mixed-state density matrix $\widehat \Gamma_{\rm imp}^{x,D,\mu} \in {\mathcal S}\left({\rm Fock}\left(\mathcal H_{\rm imp}^{x,D}\right)\right)$, satisfying the representability conditions $\widehat \Gamma_{\rm imp}^{x,D}  \ge 0$ and $\tr\left(\widehat \Gamma_{\rm imp}^{x,D} \right)=1$. In this state, the number of particles in the impurity is $\tr\left(\widehat N_{\rm imp}^{x,D}  \widehat \Gamma_{\rm imp}^{x,D} \right)$, while the number of particles in the fragment is $\tr\left(  \widehat N^{x,D}_{\rm frag} \widehat  \Gamma_{\rm imp}^{x,D}\right)$.  We want to enforce two conditions: (i) for each impurity problem, the total number of electrons in the system ``impurity plus environment'' should be $N$, (ii) when all the solutions of the $N_{\rm f}^D$ impurity problems are combined, the total number of electrons in the $N_{\rm f}^D$ fragments should be equal to $N$. Conditions (i) and (ii) respectively read
\begin{align}\label{eq:charge_constraints}
&\tr\left(\widehat N_{\rm imp}^{x,D}  \widehat \Gamma_{\rm imp}^{x,D} \right)=N-N^{x,D}_{\rm env}, \\
&\sum_{x=1}^{N_{\rm f}^D} 
\tr\left(\widehat N_{\rm frag}^{x,D}  \widehat \Gamma_{\rm imp}^{x,D} \right)=N, \label{eq:charge_constraints_2}
\end{align}
where $N^{x,D}_{\rm env}:=\tr(P^{x,D}_{\rm env}D)$ is the number of particles in the environment space $\mathcal H^{x,D}_{\rm env}$ in a state with 1-RDM $D$, with $P^{x,D}_{\rm env}$ the orthogonal projector from $\mathcal{H}$ onto $\mathcal H^{x,D}_{\rm env}$. This leads us to consider the following optimization problem for each impurity
\begin{equation}
\label{eq:def_FHL1}
\widehat \Gamma_{\rm imp}^{x,D,\mu} 
:= \hspace{-5mm}
\mathop{\rm argmin}_{\substack{\widehat \Gamma_{\rm imp}^{x,D} \in {\mathcal S}\left({\rm Fock}\left(\mathcal H_{\rm imp}^{x,D}\right)\right),  \\ \widehat \Gamma_{\rm imp}^{x,D}  \ge 0, \; \tr\left(\widehat \Gamma_{\rm imp}^{x,D} \right)=1, \\ \tr\left(\widehat N_{\rm imp}^{x,D}  \widehat \Gamma_{\rm imp}^{x,D} \right)=N-N^{x,D}_{\rm env}}}  \hspace{-5mm}
\tr\left(  \left(\widehat H_{\rm imp}^{x,D} - \mu \widehat N^{x,D}_{\rm frag} \right) \widehat  \Gamma_{\rm imp}^{x,D}\right), 
\end{equation}
where the constraint \eqref{eq:charge_constraints} is taken into account explicitly in each impurity problem, while the constraint \eqref{eq:charge_constraints_2} is satisfied by introducing a global chemical potential $\mu$. 
Next, we denote
$$
D^{xx,D,\mu}_{\rm HL} := P^{x,D}_{\rm imp \to frag} D_{\rm imp}^{x,D,\mu} \left(P^{x,D}_{\rm imp \to frag}\right)^*,
$$
where $P^{x,D}_{\rm imp \to frag}: \mathcal H^{x,D}_{\rm imp} \to \mathcal H^{x,D}_{\rm frag}$ is the orthogonal projector from the impurity space to the fragment space, and $D_{\rm imp}^{x,D,\mu}$ is the 1-RDM associated with $\widehat \Gamma_{\rm imp}^{x,D,\mu}$. The chemical potential $\mu^{D} \in \R$ is chosen such that
$$
\sum_{x=1}^{N_{\rm f}^D} 
\tr\left(\widehat N_{\rm frag}^{x,D}  \widehat \Gamma_{\rm imp}^{x,D,\mu^D} \right)=\sum_{x =1}^{N_{\rm f}^D} \tr\left( D^{xx,D,\mu^D}_{\rm HL} \right) = N,
$$
and we define
\begin{align} 
&D^{xx,D}_{\rm out}:=D^{xx,D,\mu^D}_{\rm HL}, \nonumber \\  & g(D):=\left(D^{11,D}_{\rm out}, \cdots, D^{N_{\rm f}^DN_{\rm f}^D,D}_{\rm out}\right). \label{eq:HL_map}
\end{align}
The optimization problems~\eqref{eq:def_FHL1} are referred to as the {\it high-level problems}, of the corresponding fragments. Similar to conventional DMET, in the proposed framework, we may encounter two issues (1) the possible non-uniqueness of the solution to~\eqref{eq:def_FHL1} and (2) the possible non-existence of a Lagrange multiplier $\mu$ fulfilling~\eqref{eq:charge_constraints_2}.

In conventional DMET, we have $N^{x}_{\rm env}=N-\ell^{x} \in \mathbb N$, so that the impurity problem~\eqref{eq:def_FHL1} can be solved in the $\ell^{x}$-particle sector of the impurity Fock space.

\subsubsection{Low-level solver}

In this final step, we construct the off-diagonal blocks of the resulting 1-RDM \(D_{\rm out}\). This can be viewed as a matrix-completion problem. To this end, we introduce the fragment diagonal-block extractor operator 
$$
{\rm Bd}^D: {\mathcal S}\left(\mathcal H\right) \to \mathcal S\left(\mathcal H^{1,D}_{\rm frag}\right) \times \cdots \mathcal S\left(H^{N_{\rm f}^D,D}_{\rm frag}\right),
$$
defined by 
$$
{\rm Bd}^D (M) = \left( P^{1,D}_{\rm frag} M  \left(P^{1,D}_{\rm frag}\right)^*, \cdots, P^{N_{\rm f}^D,D}_{\rm frag} M  \left(P^{N_{\rm f}^D,D}_{\rm frag}\right)^* \right).    
$$
A generic low-level solver can be written as
$$
D_{\rm out} = \mathop{\rm argmin}_{D \in \mathcal D_{\rm DMET} \; | \; {\rm Bd}^{D_{\rm in}}(D)=g(D_{\rm in})} E(D),
$$
where $E: \mathcal D_{\rm DMET}  \to \R$ is a given objective function and ${\rm Bd}^{D_{\rm in}}(D)=g(D_{\rm in})$ are linear constraints. In conventional DMET, $E$ is chosen to be the Hartree-Fock energy functional associated with $\widehat H$.

To summarize, conventional DMET has the following features:

(a) The low-level solver minimizes the Hartree-Fock energy functional over the set of orthogonal projectors with fixed diagonal blocks;

(b) The 1-RDM encoding a low-level description of the global quantum state is an orthogonal projector, which implies that there exists a canonical way to construct the impurity subspaces and the corresponding Hamiltonians;

(c) The 1-RDM fragmentation is commonly fixed by the user based on domain knowledge and remains unchanged throughout the self-consistent field (SCF) iteration.

\medskip

As concluded  in~\cite{faulstich2022pure}, the conventional DMET method faces a ``trilemma'' involving three competing conditions: (1) achieving exact fitting, that is, the $N$-representability of the target density matrix highlighted in~\cite[Lemma~9]{CFKLL2025} (i.e. ${\rm Bd}^{D_{\rm in}}(D_{\rm out})=g(D_{\rm in})$), (2) yielding an idempotent density matrix, and (3) adhering to the Aufbau principle, i.e., the low-level 1-RDM is the spectral projector onto the space spanned by the lowest $N$ orbitals of the low-level Hamiltonian. Ref.~\cite{faulstich2022pure} showed that there exist scenarios when these conditions cannot be simultaneously satisfied. Although different ways of relaxing these conditions can provide potential pathways for generalizing and improving DMET, no generally satisfactory framework has been proposed yet. Notably, relaxing condition (2) from an idempotent to a generic density matrix, as in the SDP-based approach of \cite{wu2020enhancing}, can lead to an uncontrollable increase in the number of additional bath orbitals

\section{New variants of DMET}

In this section, we introduce a generalization of DMET that overcomes the limitations discussed in the previous section; see~Fig.~\ref{fig:dmet_workflow} for an illustrated workflow. The generalized approach presented here increases the number of bath orbitals in a controlled manner through disentanglement. It differs from both conventional DMET and the most previous extensions of the theory in three key ways:

\begin{figure*}
    \centering
    \includegraphics[width=1.\textwidth]{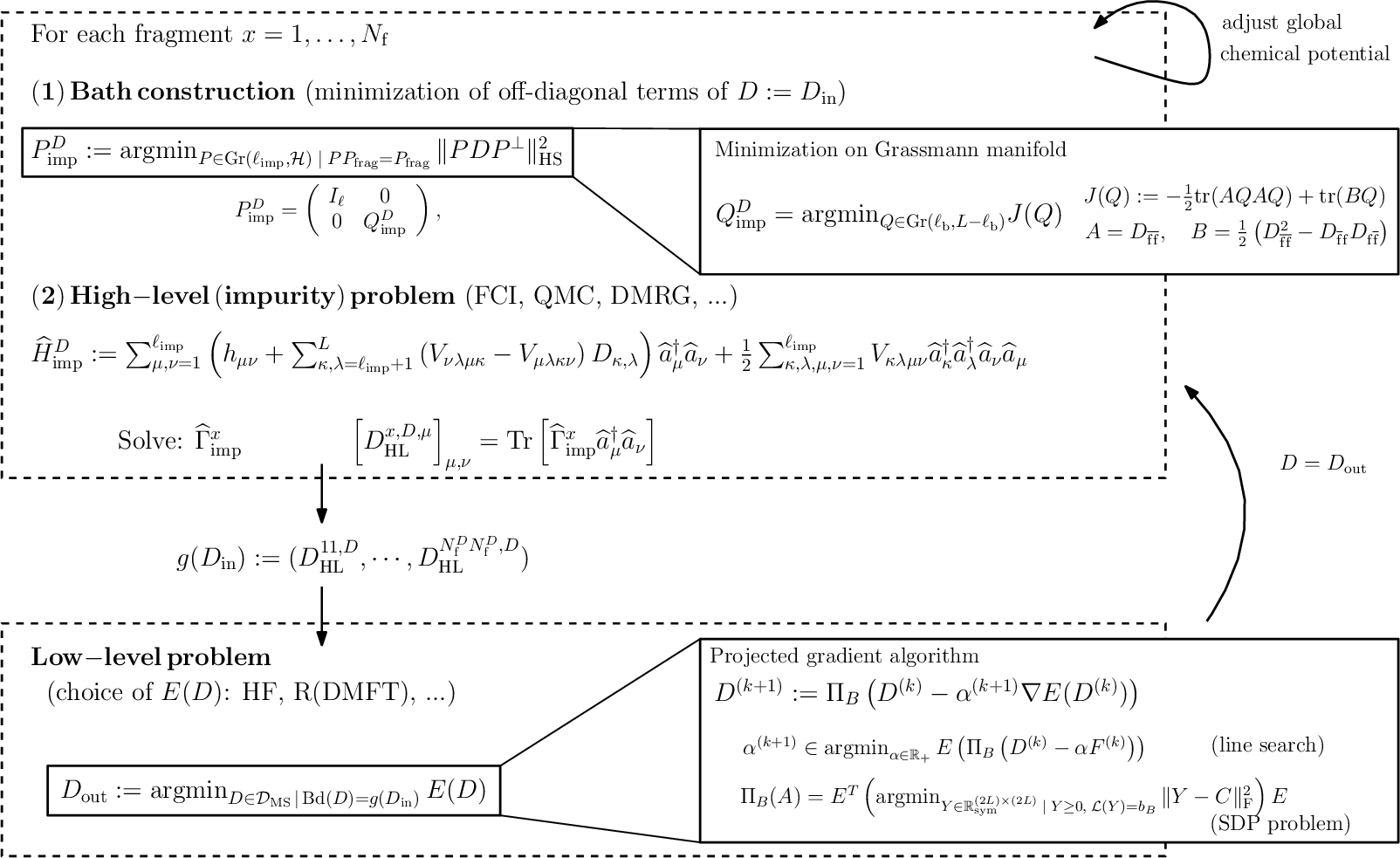}
    \caption{\textit{Graphical summary of a generalized DMET workflow}  }
    \label{fig:dmet_workflow}
\end{figure*}

(i) First, global quantum states are described by fully generic 1-RDMs able to represent any $N$-electron pure or mixed state. This flexibility allows one to consider a broader class of low-level models, for instance, objective functions given by approximations of the exact Reduced Density-Matrix Functional, as already suggested in~\cite{Sekaran2022}, see Subsection~\ref{sec:low-level}.

(ii) Second, our bath construction method is more flexible and provides a systematic way to construct a bath of arbitrarily chosen dimension for any fragment space and any low-level 1-RDM (see Subsection~\ref{sec:bath_construction}). In particular, our method does not suffer from the $N$-representability problem highlighted in~\cite[Lemma~9]{CFKLL2025}, and paves the way to black-box adaptive fragmentation methods (see Subsection~\ref{sec:fragmentation}).

(iii) Third, an ansatz must be made to construct the impurity Hamiltonian (see Subsection~\ref{sec:downfolding}). In conventional DMET, the low-level density matrix is an orthogonal projector, uniquely specifying the underlying $N$-particle quantum state. In our approach, however, the low-level density matrix is not an orthogonal projector and thus does not fully determine the state of the $N$-particle---thus requiring an ansatz.
In addition, the so-defined impurities generally do not contain an integer number of particles, so each impurity problem must be solved in the corresponding impurity Fock space and cannot be reduced to a many-body problem set in a single $n$-body sector of the impurity Fock space.

\subsection{New Bath Construction Method}
\label{sec:bath_construction}

We propose a bath construction method designed to maximally disentangle
the impurity from its environment. Specifically, the objective is to make the 1-RDM $D$ as close as possible to a block-diagonal form. To this end, consider a density matrix $D \in \mathcal{D}_{\rm MS}$, a fragment subspace $\mathcal{H}_{\rm frag} \subset \mathcal{H}$ with $\dim(\mathcal{H}_{\rm frag}) = \ell$, and a bath dimension $\ell_{\rm b}$. Our goal is to construct a bath space $\mathcal{H}_{\rm bath}^D \subset \mathcal{H}_{\rm frag}^\perp$ of dimension $\ell_{\rm b}$. Equivalently, this defines an impurity subspace $\mathcal{H}_{\rm imp}^D = \mathcal H_{\rm frag} \oplus \mathcal H^D_{\rm bath} \subset \mathcal{H}$ of dimension $\ell_{\rm imp} := \ell + \ell_{\rm b}$, which contains both the fragment space $\mathcal{H}_{\rm frag}$ and the corresponding bath space $\mathcal{H}_{\rm bath}^D$. The proposed method minimizes the magnitude of the off-diagonal blocks of the matrix $D$ in the decomposition $\mathcal{H} = \mathcal{H}_{\rm imp}^D \oplus \mathcal{H}_{\rm env}^D$, where $\mathcal{H}_{\rm env}^D := (\mathcal{H}_{\rm imp}^D)^\perp$. This ensures that the coupling between the impurity and the environment is as small as possible, resulting in a near block-diagonal form for $D$. For readability, we have simplified the notation by omitting the fragment index $x$. The fragment-specific quantities presented here apply to any arbitrary fragment.

\medskip

We choose the Hilbert--Schmidt norm, i.e. $\|A\|_{\rm HS}^2=\tr(A^*A)$, to measure the magnitude of the impurity blocks in $D$, although other choices involving the Hamiltonian $\widehat H$ could also be considered. The corresponding orthogonal projector $P_{\rm imp}^D\in {\rm Gr}(\ell_{\rm imp},\mathcal H)$ onto $\mathcal H^D_{\rm imp}$ is given by
\begin{equation}\label{eq:new_imp_proj}
P_{\rm imp}^{D} := \mathop{\rm argmin}_{P \in {\rm Gr}(\ell_{\rm imp},\mathcal H) \; | \; PP_{\rm frag}=P_{\rm frag}} \| P D P^\perp \|_{\rm HS}^2.
\end{equation} 

We emphasize that the impurity space constructed using this method is equivalent to the impurity space in conventional DMET when $\ell_{\rm b}=\ell$ and $D$ is an orthogonal projector. In this specific case, the space ${\mathcal H}_\star^D:={\mathcal H}_{\rm frag} \oplus D{\mathcal H}_{\rm frag}$ (which, by construction, contains $\mathcal H_{\rm frag}$) is invariant under $D$; similarly, the orthogonal complement $\left({\mathcal H}_\star^D\right)^\perp$ is $D$-invariant since $D$ is symmetric.
Consequently, the off-diagonal blocks of $D$ in the decomposition ${\mathcal H}_\star^D \oplus \left({\mathcal H}_\star^D\right)^\perp$ vanish. Moreover, if ${\rm dim}\left({\mathcal H}_\star^D\right)=2\ell$---which is typically the case---then $\mathcal H^D_{\rm imp}=\mathcal H_\star^D$, and we recover the impurity space used in conventional DMET. In the degenerate case $\ell_*:={\rm dim}\left({\mathcal H}_\star^D\right) < 2\ell$, the bath construction problem has infinitely many solutions: any ${\mathcal H}_{\rm imp}^D$ of the form ${\mathcal H}_{\rm imp}^D={\mathcal H}_\star^D \oplus \widetilde {\mathcal H}$, where $\widetilde {\mathcal H}$ is any subspace of ${\mathcal H}$ spanned by a set of $2\ell-\ell_*$ orthonormal eigenvectors of $D$, is an admissible impurity space.
Note also that in the case when $\ell_{\rm b}=\ell$, but $D$ is not an orthogonal projector, the bath construction method based on the Householder transformation~\cite{Sekaran2021, yalouz2022quantum} simply amounts to choosing $\mathcal H_{\rm imp}=\mathcal H_{\rm frag}+D\mathcal H_{\rm frag}$.

\medskip

For a practical instance of the proposed variants of DMET, we aim to solve~\eqref{eq:new_imp_proj} in the general case. To that end, we consider an orthonormal basis of $(\chi_1,\cdots,\chi_L)$ of $\mathcal H$ such that ${\mathcal H}_{\rm frag}={\rm Span}(\chi_1,\cdots,\chi_\ell)$ and identify ${\mathcal H} \simeq \R^L$. In this basis, $D$, $P_{\rm frag}$ and $P_{\rm imp}^D$ are represented by block matrices
$$
\left( \begin{array}{cc} D_{\rm ff} & D_{\rm f \overline{f}} \\ D_{\rm \overline{f}f} & D_{\rm \overline{f}\overline{f}} \end{array} \right), \qquad \left( \begin{array}{cc} I_\ell & 0  \\ 0 & 0 \end{array} \right), \qquad 
\left( \begin{array}{cc} I_{\ell} & 0  \\ 0 & Q_{\rm imp}^{D} \end{array} \right),
$$
where $D_{\rm ff} \in \R^{\ell \times \ell}$, $D_{\rm f\overline{f}} \in \mathbb{R}^{\ell \times (L-\ell)}$, $D_{\rm \overline{f}f} = D_{\rm f\overline{f}}^T \in \mathbb{R}^{(L-\ell) \times \ell}$, $D_{\rm \overline{f}\overline{f}},Q_{\rm imp}^{D} \in \mathbb{R}^{(L-\ell) \times (L-\ell)}$, respectively.
In the general case, where $D$ is a generic element of ${\mathcal D}_{\rm MS}$ and $1\leq \ell_{\rm b} \leq L-\ell-1$ arbitrarily, the matrix $Q_{\rm imp}^{D}$ is determined by solving the following Grassmann quadratic optimization problem 
\begin{align} 
\label{eq:Grassmann_quadratic}
\min_{Q \in {\rm Gr}(m,\R^M)} J(Q), 
\end{align}
with $m=\ell_{\rm b}$, $M=L-\ell$, and
$$
J(Q):=- \frac 12 \tr(AQAQ) + \tr(BQ)
$$
where
$$
A=D_{\rm \overline{f}\overline{f}}\in \R^{M \times M}_{\rm sym}, \quad B=\frac 12 \left(D_{\rm \overline{f}\overline{f}}^2-D_{\rm \overline{f}f}D_{\rm f\overline{f}}\right)\in \R^{M \times M}_{\rm sym}.
$$
The minimization problem~\eqref{eq:Grassmann_quadratic} can be solved either self-consistently, by solving the Euler-Lagrange equations, or by direct minimization methods.
A good initialization for either procedure is the orthogonal projector onto the $m$ eigenvectors of $A$ with eigenvalues closest to $1/2$.
It can be verified that this initialization procedure actually provides the exact solution to~\eqref{eq:Grassmann_quadratic} when $\ell_{\rm b} = \ell$ and $D$ is an orthogonal projector. When $\ell_{\rm b} = \ell$ but $D$ is not an orthogonal projector, an alternative initialization based on the Householder transformation~\cite{Sekaran2021, yalouz2022quantum} can be employed.

Let us now choose the basis vectors $(\chi_1,\cdots,\chi_L)$ so that ${\mathcal H}_{\rm frag}={\rm Span}(\chi_1,\cdots,\chi_\ell)$ and ${\mathcal H}_{\rm bath}^D={\rm Span}(\chi_{\ell+1},\cdots,\chi_{\ell+\ell_{\rm b}})$, and identify $\mathcal H \simeq \R^L$.
The 1-RDM $D$ is then represented by the block matrix (also denoted by $D$ for simplicity)
\begin{align} 
\label{eq:dec_D_imp_env}
D=\left( \begin{array}{cc}  D_{\rm ii} &  D_{\rm ie}  \\ D_{\rm ei} & D_{\rm ee} \end{array} \right), 
\end{align}
$$
D_{\rm ii} \in \R^{\ell_{\rm imp} \times \ell_{\rm imp}}_{\rm sym}, \quad D_{\rm ee} \in \R^{(L-\ell_{\rm imp}) \times (L-\ell_{\rm imp})}_{\rm sym},
$$
with $\ell_{\rm imp} = \ell + \ell_{\rm b}$.
If $D$ is an orthogonal projector as in conventional DMET, then $D_{\rm ie}=D_{\rm ei}^T=0$, $\tr(D_{\rm ii})=\ell$ and $\tr(D_{\rm ee})=N-\ell$. In the general case, $D_{\rm ie}=D_{\rm ei}^T$ is not zero but our bath construction method makes it as small as possible (for the Hilbert--Schmidt norm chosen here as a matter of example). The major difference with the orthogonal projector case is that $\tr(D_{\rm ii})$ and $\tr(D_{\rm ee})=N-\tr(D_{\rm ii})$ are not necessarily integers. As a consequence, the high-level impurity problems~\eqref{eq:def_FHL1} have to be solved in the Fock space. An alternative is to impose that the impurity (and thus the environment) contains an integer number of electrons. This can be achieved exactly by adding the linear constraint $\tr(PD)=\ell'$, where $\ell'$ is, for example, the closest integer to $\tr\left(P_{\rm imp}^DD\right)$ with $P_{\rm imp}^D$ being the unconstrained solution, or approximately by penalizing the deviations from $\tr(PD)=\ell'$.

\subsection{Downfolding: construction of the impurity model}
\label{sec:downfolding}

Let $D \in {\mathcal D}_{\rm MS}$ and ${\mathcal H}_{\rm imp}^{x,D}$ the impurity subspace built with the scheme presented in the previous subsection (\ref{sec:bath_construction}).
Subsequently, we drop the superscript $x$, and sometimes the superscript $D$ to lighten the notation. We propose a general downfolding method that coincides with the downfolding used in conventional DMET when $D$ is an orthogonal projector and ${\mathcal H}_{\rm imp}^D:={\mathcal H}_{\rm frag} \oplus D {\mathcal H}_{\rm frag}$. To that end, we first consider the approximation 
$$
D_{\rm app}=\left( \begin{array}{cc}  D_{\rm ii} & 0 \\ 0 &  D_{\rm ee}\end{array} \right),
$$
where the block structure is defined with respect to the decomposition ${\mathcal H} = {\mathcal H}_{\rm imp}^D \oplus {\mathcal H}_{\rm env}^D$ (cf.~\eqref{eq:dec_D_imp_env}).
This approximation is meaningful as the impurity space $\mathcal H_{\rm imp}^{D}$ is precisely chosen to minimize the Frobenius norm of the off-diagonal blocks of $D$.
Since $D_{\rm app}\in \mathcal D_{\rm MS}$, it is an admissible 1-RDM for the global system, and its diagonal blocks $D_{\rm ii}$ and $D_{\rm ee}$ are admissible 1-RDMs corresponding to mixed states in ${\rm Fock}\left(\mathcal H_{\rm imp}^D\right)$ and ${\rm Fock}\left(\mathcal H_{\rm env}^D\right)$, respectively. Moreover, consider {\em any} particle-number conserving many-body density matrices
$$
\widehat \Gamma_{\rm imp}^D \in \mathcal S\left({\rm Fock}\left(\mathcal H_{\rm imp}^D\right)\right) \quad \mbox{and} \quad \widehat \Gamma_{\rm env}^D \in \mathcal S\left({\rm Fock}\left(\mathcal H_{\rm env}^D\right)\right)
$$
such that the 1-RDM of the factored state $\widehat \Gamma^D:=\widehat \Gamma_{\rm imp}^D \otimes \widehat \Gamma_{\rm env}^D$ equals $D_{\rm app}$. If $D$ is an orthogonal projector and $\ell_{\rm imp}=2\ell$, then (in the non-degenerate case) $D_{\rm app}=D$ and $\Gamma^D = |\Psi^D\rangle\langle \Psi^D|$, where $\Psi^D$ is the unique Slater determinant with 1-RDM~$D$. 

By analogy with \eqref{eq:Himp_st}, we thus define the impurity Hamiltonian in the general case as follows: for any particle-number conserving $\widehat \Gamma_{\rm imp} \in \mathcal S\left({\rm Fock}\left(\mathcal H_{\rm imp}^D\right)\right)$,
\begin{equation}
\begin{aligned}
\label{eq:def_impurity_Hamiltonian}
\tr &\left( \widehat H \left(  \widehat \Gamma_{\rm imp}  \otimes  \widehat \Gamma_{\rm env}^0  \right)  \right)\\ 
&=  
\tr\left( \widehat H_{\rm imp}^D \widehat \Gamma_{\rm imp} \right)  
+
E_{\rm env}^0 \tr\left( \widehat \Gamma_{\rm imp} \right) ,
\end{aligned}
\end{equation} 
with 
\begin{widetext}
\begin{align}
\widehat H_{\rm imp}^D :&= \sum_{\mu,\nu=1}^{\ell_{\rm imp}} \left( h_{\mu\nu} + \sum_{\kappa,\lambda=\ell_{\rm imp}+1}^L \left( V_{\nu\lambda\mu\kappa} -  V_{\mu\lambda\kappa\nu} \right)D_{\kappa\lambda} \right) \widehat a_\mu^\dagger \widehat a_\nu + \frac 12
\sum_{\kappa,\lambda,\mu,\nu=1}^{\ell_{\rm imp}}  V_{\kappa\lambda\mu\nu} \widehat a_\kappa^\dagger  \widehat a_\lambda^\dagger \widehat a_\nu \widehat a_\mu,  \label{eq:impurity_Hamilotian_expression} \\
 E_{\rm env}^0 :&=\sum_{\mu,\nu=\ell_{\rm imp}+1}^L h_{\mu\nu} D_{\mu\nu} + \frac 12 \sum_{\kappa,\lambda,\mu,\nu=\ell_{\rm imp}+1}^L V_{\kappa\lambda\mu\nu} [\Gamma_{\rm env}^0]_{\kappa\lambda\nu\mu}. \label{eq:env_energy}
\end{align}
\end{widetext}
We emphasize that in the derivation of~\eqref{eq:def_impurity_Hamiltonian}--\eqref{eq:env_energy},
we used that both $\widehat \Gamma_{\rm imp}$ and $\widehat\Gamma_{\rm env}^0$ are particle-number conserving. Therefore, all two-body terms with one index smaller or equal to $\ell_{\rm imp}$ and three indices larger than $\ell_{\rm imp}$---or vice versa---vanish.

\medskip

It remains to solve the high-level problem~\eqref{eq:def_FHL1}. Dropping the superscript $x$, and defining 
$$
\widehat K^{D,\mu}_{\rm imp} :=\widehat H_{\rm imp}^{D} - \mu \widehat N_{\rm frag}, \quad N^D_{\rm imp}:=N-N^D_{\rm env},
$$
problem~\eqref{eq:def_FHL1} can be rewritten as
\begin{equation}
\label{eq:def_FHL11}
\widehat \Gamma_{\rm imp}^{D,\mu} = \widehat \Gamma^{D,\mu}\left[N_{\rm imp}^D\right],
\end{equation}
where for any \emph{real} number $0 \le n \le {\rm dim}\left(\mathcal H_{\rm imp}^{D}\right)$, with $n$ representing the particle number, which may be non-integer,
\begin{equation}
\label{eq:def_FHL12}
\Gamma^{D,\mu}[n] := \hspace{-5mm}
\mathop{\rm argmin}_{\substack{\widehat \Gamma_{\rm imp}^{D} \in {\mathcal S}\left({\rm Fock}\left(\mathcal H_{\rm imp}^{D}\right)\right),  \\ \widehat \Gamma_{\rm imp}^{D}  \ge 0, \; \tr\left(\widehat \Gamma_{\rm imp}^{D} \right)=1, \\ \tr\left(\widehat N_{\rm imp}^{D}  \widehat \Gamma_{\rm imp}^{D} \right)=n}}  \hspace{-5mm}
\tr\left(  \widehat K^{D,\mu}_{\rm imp} \widehat  \Gamma_{\rm imp}^{D}\right).
\end{equation}
The main difference to conventional DMET is that $N^D_{\rm imp}$ is not necessarily an integer. For $M \in [\![{\rm dim}\left(\mathcal H^D_{\rm imp}\right)]\!]$, where we use the notation $[\![\ell ]\!] = \{1,2,...,\ell\}$,  let $E^{D,\mu,0}_{{\rm imp},M}$ denote the ground-state energy of $\widehat K^{D,\mu}_{\rm imp}$ in the $M$-particle sector of ${\rm Fock}\left(\mathcal H_{\rm imp}^{D}\right)$ and $\Psi^{D,\mu,0}_{{\rm imp},M} \in \bigwedge^M \mathcal H_{\rm imp}^{D}$ is an associated normalized ground-state wavefunction. Since the grand-canonical Hamiltonian $\widehat K^{D,\mu}_{\rm imp}$ is particle-number conserving, the function 
\begin{equation}
\label{eq:def_FHL13}
\mathcal E^{D,\mu}(n) := \hspace{-5mm}
\mathop{\rm min}_{\substack{\widehat \Gamma_{\rm imp}^{D} \in {\mathcal S}\left({\rm Fock}\left(\mathcal H_{\rm imp}^{D}\right)\right),  \\ \widehat \Gamma_{\rm imp}^{D}  \ge 0, \; \tr\left(\widehat \Gamma_{\rm imp}^{D} \right)=1, \\ \tr\left(\widehat N_{\rm imp}^{D}  \widehat \Gamma_{\rm imp}^{D} \right)=n}}  \hspace{-5mm}
\tr\left(  \widehat K^{D,\mu}_{\rm imp} \widehat  \Gamma_{\rm imp}^{D}\right)
\end{equation}
is the convex hull of the discrete function 
$$
[\![{\rm dim}(\mathcal H_{\rm imp}^{D})]\!] \ni M  \mapsto E^{D,\mu,0}_{{\rm imp},M} \in \R.
$$
In principle, it would be necessary to compute all ${\rm dim}(\mathcal H_{\rm imp}^{D})+1$ values $E^{D,\mu,0}_{{\rm imp},M}$ to obtain the solution to \eqref{eq:def_FHL12}. In practice, it is reasonable to make the assumption that if $N_{\rm imp}^D=(1-t) M + t (M+1)$ with $M:=\lfloor N_{\rm imp}^D \rfloor$ and $t \in [0,1)$, then 
\begin{equation}\label{eq:conv_comb}
\mathcal E^{D,\mu}(N_{\rm imp}^D)=(1-t) E^{D,\mu,0}_{{\rm imp},M} + t E^{D,\mu,0}_{{\rm imp},M+1}.
\end{equation}
This is in particular the case if 
$$
2 E^{D,\mu,0}_{{\rm imp},M} < E^{D,\mu,0}_{{\rm imp},M-1}+E^{D,\mu,0}_{{\rm imp},M+1}
$$
for all $M\in[\![{\rm dim}(\mathcal H_{\rm imp}^{D})-1]\!]$.
This stability condition is equivalent to the condition that ionization energies are higher than electron affinities. Under the assumption~\eqref{eq:conv_comb}, a solution to~\eqref{eq:def_FHL12} with $n=N^D_{\rm imp}$ is given by 
$$
\widehat \Gamma_{\rm imp}^{D,\mu} =(1-t) |\Psi^{D,\mu,0}_{{\rm imp},M}\rangle\langle\Psi^{D,\mu,0}_{{\rm imp},M}| + t |\Psi^{D,\mu,0}_{{\rm imp},M+1}\rangle \langle \Psi^{D,\mu,0}_{{\rm imp},M+1}|.
$$

\subsection{Low-level solvers}
\label{sec:low-level}

Let $D_{\rm in} \in \mathcal D_{\rm MS}$ and let $g(D_{\rm in})$ denote the collection of the diagonal blocks of the output matrix $D_{\rm out}$ in the fragment decomposition $ \mathcal H^{1,D_{\rm in}}_{\rm frag} \oplus \cdots \oplus \mathcal H^{N_{\rm f}^{D_{\rm in}},D_{\rm in}}_{\rm frag}$ of $\mathcal H$, as defined in  \eqref{eq:HL_map}. 
Choose an orthonormal basis $(\chi_1,\cdots,\chi_L)$ of $\mathcal H$ that is compatible with this decomposition, so that we can identify $\mathcal H$ with~$\R^L$. In this basis, the map ${\rm Bd}:={\rm Bd}^{D_{\rm in}}$ simply extracts the diagonal blocks of a real symmetric matrix, with block sizes $\left(\ell^{1,D_{\rm in}},\cdots,\ell^{N_{\rm f}^{D_{\rm in}},D_{\rm in}}\right)$.

\medskip

In the above matrix representation, the conventional DMET low-level solver is given by the constrained Hartree-Fock problem
$$
D_{\rm out}:=\mathop{\rm argmin}_{D \in {\mathcal D}_{\rm Slater} \, | \, {\rm Bd}(D)=g(D_{\rm in})} E^{\rm HF}(D),
$$
where $E^{\rm HF}$ is the Hartree-Fock energy functional.

\medskip

Since we now work within the set ${\mathcal D}_{\rm MS}$ of all mixed-state $N$-particle 1-RDMs, we can consider more general low-level solvers of the form
\begin{equation}\label{eq:def_FLL_new}
D_{\rm out} :=\mathop{\rm argmin}_{D \in {\mathcal D}_{\rm MS} \, | \, {\rm Bd}(D)=g(D_{\rm in})} E(D),
\end{equation}
where $E$ is any approximation of the exact 1-RDM energy functional in the chosen basis. The gapless problem in conventional DMET~\cite{wu2020enhancing} arises because the minimization is performed over $\mathcal{D}_{\rm Slater}$ rather than $\mathcal{D}_{\rm MS}$. Since our formulation does not impose this restriction, the gapless issue does not arise here. Moreover, since $\mathcal{D}_{\rm MS}$ is a compact and convex set, and the map $D \mapsto {\rm Bd}(D)$ adds only linear constraints, the feasible domain in \eqref{eq:def_FLL_new} remains closed and convex. If the approximated energy functional $E$ is continuous (or convex), then a global minimizer exists, which guarantees the solvability of \eqref{eq:def_FLL_new}. In cases where multiple minimizers arise, adding a small regularization term (e.g. entropy-based penalty) can help ensure uniqueness or guide the solution toward a more physically preferred one.
Recall that (R)DMFT is {\em in principle} an exact theory, just as DFT. Proceeding similarly as in~\cite{Valone80,Lieb83}, the exact $N$-particle (mixed) ground-states 1-RDM of $\widehat H$ can be obtained by the constrained search method
$$
\min_{D \in {\mathcal D}_{\rm MS}} E^{\rm exact}(D) \quad  \mbox{where} \quad E^{\rm exact}(D) := \min_{\widehat \Gamma \mapsto D} \tr\left( \widehat H \widehat \Gamma \right).
$$
The functional $E^{\rm exact}(D)$ can be decomposed as 
$
E^{\rm exact}(D) = E^{\rm HF}(D) + E_{\rm c}(D)$ where
\begin{align*}
    E_{\rm c}(D):=&\min_{\widehat \Gamma \mapsto D} \tr\left( \widehat H_{\rm 2-body} \widehat \Gamma \right) \\ & - \left( \frac 12 \tr(J(D)D) - \frac 12 \tr(K(D)D) \right)
\end{align*}
is the 1-RDM correlation energy functional. As in DFT, no simple expression of the functional $E_{\rm c}$ is known, so approximations must be used. The simplest approximation consisting in neglecting $E_{\rm c}$ gives back the Hartree--Fock model. Constructing accurate approximations of $E_{\rm c}$ is even more challenging than constructing accurate approximations of the exchange-correlation energy functional in DFT. In this perspective, DMET can be interpreted as an approximate (R)DMFT in which the limitations of the approximate 1-RDM energy functional $E$ are partly compensated by the fact that $E$ is minimized over a set of 1-RDMs whose diagonal blocks contain some high-level correlation.

Let us finally present an algorithm to solve \eqref{eq:def_FLL_new}. To simplify the notation, we set $B:=g(D_{\rm in})$ and 
 denote by $\Pi_B$ the orthogonal projector (with respect to the Frobenius inner product) on the non-empty closed convex set 
$$
\mathcal K_B = \{ D \in \mathcal D_{\rm MS} \; | \; {\rm Bd}(D)=B \}.
$$
Note that computing $\Pi_B(A)$ for $A \in \R^{L \times L}_{\rm sym}$ amounts to solving a semidefinite least square problem. Indeed, we have
\begin{align*}
\Pi_B(A)&=\mathop{\rm argmin}_{D \in \mathcal K_B} \|D-A\|^2,
\end{align*}
and the above problem can be reformulated as 
$$
\Pi_B(A) = E^T \left(  \mathop{\rm argmin}_{Y \in \R^{(2L) \times (2L)}_{\rm sym} \; | \; Y \ge 0, \; \mathcal{L}(Y) = b_B} \|Y-C\|_{\rm F}^2 \right) E,
$$
where
$$
C := \left( \begin{array}{cc} A & 0 \\ 0 & I_L-A \end{array} \right), \; 
Y := \left( \begin{array}{cc} D & 0 \\ 0 & I_L-D \end{array} \right) \in \R^{(2L) \times (2L)}_{\rm sym} 
$$
and
$$
E:=  \left( \begin{array}{c} I_L \\ 0 \end{array} \right) \in \R^{(2L) \times L}.
$$
Here, the compact notation $\mathcal{L}(Y)=b_B$ collects the following affine constraints:
\begin{align*}
& Y_{L_x^<+\mu,L_x^<+\nu}=[B^x]_{\mu\nu}, \quad \forall x \in [\![N_{\rm f}^{D_{\rm in}}]\!] , \; \mu,\nu \in [\![\ell^{x,D_{\rm in}}]\!], \\
& Y_{\mu,L+\nu}=0, \quad \forall \mu,\nu  \in [\![L]\!], \\
& Y_{\mu,\nu} + Y_{L+\mu,L+\nu} = \delta_{\mu\nu}, \quad \forall \mu,\nu  \in [\![L]\!],
\end{align*}
where $L_x^< := \sum_{x'=1}^{x-1} \ell^{x',D_{\rm in}}$.

\medskip

\noindent
{\bf Projected gradient algorithm to solve the low-level problem \eqref{eq:def_FLL_new}:} 
\medskip

\noindent
Let $B:=g(D_{\rm in})$ and let $\eta_{\rm LL} > 0$ be a convergence threshold. Set $k=0$, $D^{(0)}:=\mbox{BlockDiag}(B^1,\cdots,B^{N_{\rm f}})$, $F^{(0)}=\nabla E(D^{(0)})$.

\medskip

\noindent
Iterate the following two steps until the convergence criterion $\left\|D^{(k)} - \Pi_{B} \left( D^{(k)}- F^{(k)} \right)\right\|_{\rm F} \le \eta_{\rm LL}$ is met:
\begin{enumerate}
\item Compute (an approximation of) 
$$
\alpha^{(k+1)} \in \mathop{\rm argmin}_{\alpha \in \R_+} E\left( \Pi_{B} \left( D^{(k)}- \alpha F^{(k)} \right) \right);
$$
\item  Update $D^{(k+1)}:= \Pi_{B} \left( D^{(k)}- \alpha^{(k+1)} F^{(k)}\right)$ and $F^{(k+1)}=\nabla E(D^{(k+1)})$. Set $k \leftarrow k+1$.
\end{enumerate}
Return $D_{\rm out}=D^{(k)}$. This simple projected-gradient algorithm may be accelerated via standard techniques such as DIIS.

\subsection{Toward adaptive fragmentation methods}
\label{sec:fragmentation}

Selecting an appropriate fragmentation for a given system is a challenging task that requires substantial domain expertise. As an additional feature, the proposed framework introduced in Section~\ref{sec:bath_construction} provides an alternative approach to potentially update the fragmentation adaptively based on optimization of a quantitative criterion.

\medskip

To illustrate this point, let us assume that we have a reasonable guess of the number $N_{\rm f}$ of fragments required to achieve the desired accuracy, along with rough estimates of the sizes $\ell^x$ and location (e.g., in the physical space, or in the energy space) of each of the $N_{\rm f}$ fragments, and of the sizes $\ell^x_{\rm imp}$ of the impurities. We anchor fragment $x$ in the chosen region by selecting   mutually orthogonal subspaces $\mathcal H^x_{\rm a}$ of dimension $\ell^x_{\rm a} < \ell^x$ and impose that $\mathcal H^x_{\rm a} \subset \mathcal H_{\rm frag}^x$. Let $n^x:=\ell^x-\ell^x_{\rm a}$ and $P^x_{\rm a}$ be the orthogonal projector on $\mathcal H^x_{\rm a}$. Given a trial global 1-RDM $D \in \mathcal D_{\rm MS}$, we seek non-overlapping fragment spaces $\mathcal H_{\rm frag}^{x,D}$ of the form
$$
\mathcal H_{\rm frag}^{x,D} = \mathcal H^x_{\rm a} \oplus 
\mathcal X^{x,D}, \quad \mbox{with} \quad \mbox{dim}\left(\mathcal X^{x,D}\right)=n^x,
$$
in such a way that the impurity and environments subspaces $\mathcal H_{\rm imp}^{x,D}$ and $\mathcal H_{\rm env}^{x,D}$ built from $D$ and the $\mathcal H_{\rm frag}^{x,D}$'s as in Section~\ref{sec:bath_construction} are as disentangled as possible at the 1-RDM level. Denote by $Q^{x,D}$ the orthogonal projector on the space $\mathcal X^{x,D}$, and define 
$\mathcal X:=\left( \mathcal H^1_{\rm a} \oplus \cdots \oplus \mathcal H^{N_{\rm f}}_{\rm a} \right)^\perp$. The set $Q^{\bullet,D}=(Q^{1,D},\cdots,Q^{N_{\rm f},D})$ is in fact an element of the flag manifold
\begin{align*}
\mathcal M:&=\mbox{Flag}(n_1,\cdots,n_{N_{\rm f}};\mathcal X)\\
:&=
\bigg\{ Q^\bullet=(Q^1,\cdots,Q^{N_{\rm f}}), \\
& \qquad \; Q^x \in {\rm Gr}(n^x,\mathcal X), \; Q^xQ^{x'}=0 \bigg\}.
\end{align*}
An optimal fragment decomposition is then obtained by solving the nested Riemannian optimization problem
$$
Q^{\bullet,D} = \min_{Q^{\bullet} \in \mathcal M} J(Q^\bullet),
$$
where
$$
J(Q^\bullet):=\sum_{x=1}^{N^{\rm f}} \min_{\substack{P^x \in {\rm Gr}(L^x_{\rm imp},\mathcal H) \\ P^x(P^x_{\rm a}+Q^x)=P^x_a+Q^x}} \left\|P^x D (P^x)^\perp \right\|^2.
$$
A good initial guess can be constructed using the subspaces $D^n\mathcal H^x_{\rm a}$ for $n=1,2,\cdots$, which should ideally be all contained in $\mathcal H^{x,D}_{\rm imp}$ so that the 1-RDM is fully disentangled (i.e. block diagonal in the decomposition $\mathcal H^{x,D}_{\rm imp}\oplus \mathcal H^{x,D}_{\rm env}$). This approach can be adapted to the case when the particle numbers in the impurities are constrained (either strictly or approximatively by penalization) to be integers.

If any of the terms $\|P^{x,D}_{\rm imp} D \left(P^{x,D}_{\rm imp}\right)^\perp \|^2$, where $P^{x,D}_{\rm imp}$ is the orthogonal projector on the optimized impurity space, exceed a predetermined threshold, this is an indication that the fragmentation is of poor quality and that the parameters $\ell^x$ or $\ell^x_{\rm imp}$ should be modified.

\medskip

\section{Concluding remarks}

In this article, we introduced a generalization of the DMET framework.
Our generalization is based on the observation that relaxing the idempotency requirement on the global 1-RDM leads to a more flexible definition of the impurity space.
In our proposal, we require only that the impurity space be as disentangled as possible from the environment, a condition enforced by minimizing the norm of the off-diagonal blocks of the global 1-RDM. This approach provides flexible control over the bath size, thereby enabling precise management of the computational cost.

This general requirement, along with the flexibility in bath size, distinguishes our approach from previous efforts to extend DMET. A comprehensive study comparing existing DMET methods with the proposed framework is subject to ongoing work.

In particular, our new variant opens several questions.
For instance, the dependence of the accuracy of the method on the bath dimensions is not obvious: it does not necessarily improve with increasing the bath dimensions (as in the example of a system made of two non-interacting subsystems). Moreover, this framework provides an alternative and systematic pathway for determining bath sizes that are optimal in the sense of disentangling the individual fragments.

\section*{Acknowledgments}
This project has received funding from the European Research Council (ERC) under the European Union's Horizon 2020 research and innovation programme (grant agreement EMC2 No 810367) and from the Simons Targeted Grant in Mathematics and Physical Sciences on Moir\'e Materials Magic (Award No. 896630, L.L., E.C.). 
Part of this work was done during the reunion of the IPAM program {\it Advancing quantum mechanics with mathematics and statistics}.

\bibliographystyle{apsrev4-2}
\bibliography{lib}

\end{document}